\documentclass[twocolumn,showpacs,preprintnumbers,amsmath,amssymb]{revtex4}

\usepackage{graphicx}
\usepackage{dcolumn}
\usepackage{bm}

\begin{document}

\preprint{}

\title{
Velocity correlations in the dense granular shear flows:\\
Effects on energy dissipation and normal stress
}

\author{Namiko Mitarai$^1$\footnote{Permanent address.}$^{,2}$
 and Hiizu Nakanishi$^1$}
\affiliation{$^1$
Department of physics, Kyushu University 33, Fukuoka 812-8581,
Japan.\\
$^2$
Niels Bohr Institute, Bledamsvej 17,
DK-2100, Copenhagen, Denmark.}

\date{\today}

\begin{abstract}
We study the effect of pre-collisional velocity correlations 
on granular shear flow by molecular dynamics simulations
of the inelastic hard sphere system.
Comparison of the simulations with the kinetic theory 
reveals that the theory overestimates both the
energy dissipation rate and the normal stress in 
the dense flow region.
We find that the relative normal velocity
of colliding particles is smaller than that expected
from random collisions, and the discrepancies
in the dissipation and the normal stress can be 
adjusted by introducing the idea of the 
collisional temperature, from 
which we conclude that the velocity
correlation neglected in the kinetic theory is 
responsible for the discrepancies.
Our analysis of the distributions of the pre-collisional 
velocity suggests that the correlation grows through
multiple inelastic collisions during the time scale of the inverse of the shear
rate. As for the shear stress, the discrepancy 
is also found in the dense region, but it 
depends strongly on the particle inelasticity.
\end{abstract}

\maketitle

\section{Introduction}\label{intro}
Granular media can flow like a fluid under a certain situation.
In the case of {\it the rapid granular flow}, 
where the density is relatively low and 
interactions are dominated by the instantaneous 
collisions, the kinetic theory of dense gases~\cite{Chapman} 
is extended to the inelastic hard spheres
to derive the constitutive relations~\cite{rapid}.
In the theory, the density correlations is taken into account to some extent
but not the velocity correlations in most of the cases.
As the flow gets denser, however, the molecular chaos assumption 
becomes questionable. In addition, 
the interactions may no longer be approximated 
by the instantaneous collisions but enduring contacts 
take place around the random closed packing fraction. 
The comprehensive granular rheology 
including the rather complicated
dense regime has not been established yet.

During the last several years, 
careful experiments
and large-scale molecular dynamics
simulations have been done on the 
dense granular flows~\cite{Pouliquen,Silbert,GDRMiDi,Mitarai}.
One of the important model systems that has been 
intensively studied is the steady flow down a slope under the gravity,
where we can control the ratio of the shear 
stress $S$ to the normal stress $N$ by 
changing the inclination angle $\theta$.
In this system, it has been found that
the packing fraction $\nu$ in the bulk of the flow
is constant and is determined solely by the inclination angle $\theta$;
in other words, $\nu$ is 
independent of the total flow hight $H$ and/or
the roughness of the slope~\cite{Silbert,Mitarai}.

This interesting feature has been qualitatively understood
by using {\it the Bagnold Scaling}~\cite{Bagnold},
which states the shear stress $S$ 
is proportional to
the square of the shear rate $\dot \gamma$:
\begin{equation}
S=m\sigma^{-1}A(\nu)\dot \gamma^2.
\label{bagnolds}
\end{equation}
Here, $m$ is the particle mass, and $\sigma$ is the 
particle diameter.
This scaling can be understood by dimensional analysis 
of the rigid granular flow, where the inverse of the shear rate 
$\dot \gamma^{-1}$ is the only time scale in the system.
This scaling applies to the normal stress $N$ also, which gives
\begin{equation}
N= m\sigma^{-1} B(\nu)\dot \gamma^2.
\label{bagnoldp}
\end{equation}
In the slope flow under gravity, the force balance 
gives $S/N=\tan\theta$. Thus we finally have
\begin{equation}
\frac{S}{N}=\frac{A(\nu)}{B (\nu)}=\tan \theta,
\label{slope}
\end{equation}
i.e., the packing fraction $\nu$ is determined
by the inclination angle $\theta$.

This dimensional analysis 
does not hold when the time scales other
than $\dot \gamma^{-1}$ come into the problem,
e.g. the time scales of the particle 
deformation~\cite{hatano},
but not only the constant density profile
but also the Bagnold scaling itself has
been found in the numerical simulations of dense 
steady flow down a slope for hard enough particles~\cite{Silbert}.

In the slope flow simulations, 
the value of the packing fraction $\nu$ has been 
shown to increase upon decreasing the inclination angle $\theta$, 
and eventually the flow stops at a finite 
angle $\theta_{\rm stop}$; 
namely, $A(\nu)/B(\nu)$ is an decreasing function of $\nu$
in the dense region~\cite{Silbert,Mitarai}.
One can interpret the transition at 
$\theta_{\rm stop}$ as
the jamming transition~\cite{Pouliquen,Silbertjamming}.

\begin{figure} 
\includegraphics[width=0.23\textwidth]{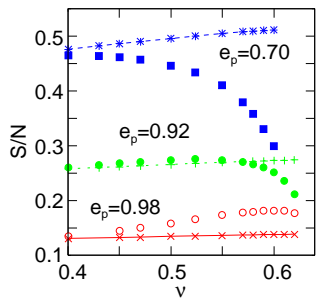}
\caption{(color online)
The ratio of the shear stress to the normal stress
$S/N$ versus the packing fraction $\nu$
from the simulation data 
($e_p=0.70$ ($\blacksquare$),
$0.92$ ($\bullet$), and $0.98$ ($\circ$))
and the plot of Eq.~(\ref{SoverP}) from the kinetic theory
($e_p=0.70$ ($*$ connected by dashed line),
$0.92$ ($+$ connected by dashed line), 
and $0.98$ ($\times$ connected by dashed line).
For the simulation data, the average normal stress 
$N=\frac{1}{3}(N_x+N_y+N_z)$ is used.
}\label{friction_packing1}
\end{figure}

The theoretical analysis of the functional form of 
$A(\nu)/B(\nu)$ has been done by Louge~\cite{Louge}
using the kinetic theory,
but he found the opposite dependence 
in the dense region, namely,  
the theory gives increasing packing fraction $\nu$
upon increasing inclination angle $\theta$
as shown in Fig.~\ref{friction_packing1},
where curves from a kinetic theory~\cite{Garzo}
is shown by symbols connected by dashed lines for 
the various restitution coefficients $e_p$.

Several explanations for this discrepancy 
have been proposed, such as 
the enduring contact~\cite{Louge,Jenkins2}, 
the Burnett order (the second order of 
the spatial gradients) effect~\cite{Kumaran}, and
the particle roughness~\cite{Kumaran} etc.,
but the subject is still under debate. 

Recently, the present authors~\cite{Mitarai} have made a detailed 
comparison between the simulation results 
of the dense slope flow and the kinetic theory
by Jenkins and Richman~\cite{Jenkins}.
In contrast with the rather good agreement for the stresses,
it has been found that the kinetic theory overestimate 
the energy dissipation rate $\Gamma$,
and this discrepancy is responsible for the contradicting behavior
in the kinetic theory,
i.e. $A(\nu)/B(\nu)$ increases with the packing fraction $\nu$.

The authors conjectured that 
the discrepancy in the energy dissipation rate
$\Gamma$  should be caused by
the velocity correlations enhanced by the inelastic collisions; 
the decrease of the relative normal velocity 
through the inelastic collisions
results in reduction of the energy loss per collision.
Such an effect has been noticed in granular gas simulations
without shear~\cite{Cooling,Bizon},
and the velocity correlations 
has been investigated analytically~\cite{Soto,Noije}.

However, the situation is rather complicated under shear, 
because the shear tends to break the correlations. 
The spatial velocity correlations 
in granular flow under shear 
has not been carefully studied so far~\cite{comment1}.

In this paper, we study the velocity correlation 
in the sheared granular flow, focusing its
effects on the energy dissipation rate
and the stress.
We adopt the simple shear flow 
of the inelastic hard spheres as a model system, 
in accordance with most of the kinetic theory analysis.
Note that
the enduring contacts is not allowed
in the hard sphere model,
whose effects are often under 
debate in the soft-sphere model simulations 
of the slope flow~\cite{Silbert,Mitarai}.

This paper is organized as follows.
In section \ref{model}, we briefly summarize 
the inelastic hard sphere model and  
the constitutive relations based on the kinetic theory.
We summarize our simulation method and present the 
results in section \ref{simulation}.
The discussion and the summary are given in section 
\ref{discussion}. 

\section{Inelastic hard sphere model and the kinetic theory}\label{model}
The inelastic hard sphere model is one of the 
simplest and widely-used models of granular 
materials~\cite{rapid,Duran}.
The particles are infinitely rigid,
and they interact through
instantaneous two-body collisions.
We adopt the simplest collision rule 
for the monodisperse smooth hard spheres with 
diameter $\sigma$, mass $m$,
and a constant normal restitution coefficient $e_p$
in three dimensions as follows:
The particle $i$ at the position $\bm r_i$ 
with the velocity $\bm c_i$ 
collides with the particle $j$ if 
$|\bm r_i-\bm r_j|=\sigma$ and 
$(\bm r_i-\bm r_j)\cdot (\bm c_i-\bm c_j)<0$,
and their post-collisional velocities 
$\bm c_i^*$ and $\bm c_j^*$  are given by
\begin{eqnarray}
\bm c_i^*&=&\bm c_i -\frac{1+e_p}{2}
\left[\bm n_{ij}\cdot 
\left(\bm c_i-\bm c_j\right)\right]\bm n_{ij}, \\
\bm c_j^*&=&\bm c_j +\frac{1+e_p}{2}
\left[\bm n_{ij}\cdot \left(\bm c_i-\bm c_j\right)\right]\bm n_{ij},
\label{ColRule}
\end{eqnarray}
respectively. Here, $\bm n_{ij}$ is a unit vector defined as
$\bm n_{ij}=(\bm r_i-\bm r_j)/|\bm r_i-\bm r_j|$.
The collision is elastic when $e_p=1$,
and inelastic when $0<e_p<1$. 
In the inelastic case, the particles lose the 
kinetic energy every time they collide,
thus external drive is necessary to 
keep particles flowing.

We compare the simulation results of the inelastic hard spheres
with the constitutive relations obtained from the
Chapman-Enskog method~\cite{Chapman}, which has been 
developed in the kinetic theory of gases. 
In this paper, we employ those by Garz\'o and
Dufty~\cite{Garzo},
who have improved the previous studies~\cite{rapid,Jenkins,Lun}, that  
is limited to the weakly inelastic case  ($(1-e_p)\ll 1$),
to include the case with any value of the restitution 
constant $e_p$ under the assumption that the state is near the local 
homogeneous cooling state~\cite{comment2}.

In the following, we briefly summarize the kinetic theory 
to derive the constitutive relations.
The hydrodynamic variables 
are the number density field $n(\bm r,t)$,
the velocity field $\bm u(\bm r,t)$,
and the granular temperature field $T(\bm r,t)$,
defined in terms of the single-particle distribution function
$f(\bm r,\bm c, t)$ as 
\begin{eqnarray}
n(\bm r,t)&=&\int 
f(\bm r,\bm c,t)\mbox d \bm c ,\\
\bm u(\bm r,t)&=&\frac{1}{n}\int \bm c f(\bm r,\bm c,t) \mbox d \bm c ,\\
T(\bm r,t)&=&\frac{m}{3n}\int (\bm c-\bm u)^2 
f(\bm r,\bm c,t) \mbox d \bm c .
\end{eqnarray}
The hydrodynamic equations for these variables are given by
\begin{eqnarray}
\frac{\partial n}{\partial t}+
\nabla \cdot (n\bm u)&=&0,\label{contituity}\\
mn\frac{\partial \bm u}{\partial t}+
mn \bm u\cdot \nabla \bm u&=&-\nabla \tensor \Sigma,\label{motion}\\
\frac{3}{2}\left(
n\frac{\partial T}{\partial t}+
n \bm u\cdot \nabla T\right)&=&-\nabla \cdot \bm q
-\tensor \Sigma:\tensor E-\Gamma,
\label{ebalance0}
\end{eqnarray}
where $\tensor \Sigma$ is the stress tensor,
$\bm q$ is the heat flux,
and $\tensor E$ is the symmetrized velocity gradient tensor:
$E_{ij}=\frac{1}{2}(\frac{\partial}{\partial r_i} u_j+
\frac{\partial}{\partial r_j} u_i)$.
Note that the energy dissipation rate $\Gamma$ 
in Eq.~(\ref{ebalance0}) appears 
due to the energy loss through the inelastic collisions,
which gives peculiar features 
to the granular hydrodynamics.

The constitutive relations for  
$\tensor \Sigma$, $\bm q$, and $\Gamma$ are determined by
the single-particle distribution $f(\bm r,\bm c, t)$. 
Its time evolution depends on the two-particle distribution function
$f^{(2)}(\bm r_1,\bm r_2,\bm c_1,\bm c_2,t)$
through the two-particle collision;
the $n$-particle distribution function 
depends on the ($n+1$)-particle distribution function.
This is known as the BBGKY hierarchy~\cite{SimpleLiquid}. 

In the Enskog approximation,
the two-particle distribution at collision
is approximated as
\begin{eqnarray}
&&
f^{(2)}(\bm r_1,\bm r_1+\sigma\bm n_{21},\bm c_1,\bm c_2,t) \nonumber\\
&=&
g_0(\nu)
f(\bm r_1,\bm c_1,t)
f(\bm r_1+\sigma\bm n_{21},\bm c_2,t) 
\label{eq:Enskog}
\end{eqnarray}
to close the BBGKY hierarchy at the 
single-particle distribution~\cite{Chapman,Garzo}.
Here, $g_0(\nu)$ is the radial distribution function
at distance $\sigma$, and depends on the packing fraction 
$\nu=\frac{1}{6}\pi \sigma^3 n$.
The term $g_0(\nu)$ represents the positional correlations,
and the actual procedure to determine the functional form of 
$g_0(\nu)$ is presented in subsection \ref{radial}.
The correlations in the particle velocities are neglected
under the molecular chaos assumption.

\begin{table}
\caption{The dimensionless functions 
in the constitutive relations from Ref.~\cite{Garzo}.
}
\label{constitutiveeq}
\begin{tabular}{cc}
\hline
$f_1(\nu)$& $\frac{6}{\pi}\nu(1+2(1+e_p)\nu g_0(\nu))$\\
$f_2(\nu)$& $
\frac{5}{16\sqrt{\pi}}
\left[\eta^{k*}\left(1+
\frac{4}{5}\nu g_0(\nu)(1+e_p)\right)
+\frac{3}{5}\gamma^*\right]
$\\
$f_3(\nu)$& $
\frac{72(1-e_p^2)}{\pi^{3/2}}\nu^2 g_0(\nu)\left(1+\frac{3}{32}c^*(e_p)\right)$\\
$\eta^{k*}$&
$
\left(\nu_\eta^*-\frac{1}{2}\zeta^{(0)*}\right)^{-1}
\left(1-\frac{2}{5}(1+e_p)(1-3e_p)\nu g_0(\nu)\right)
$\\
$\nu_\eta^*$&
$g_0(\nu)
(1-\frac{1}{4}(1-e_p)^2)(1-\frac{1}{64}c^*(e_p))$\\
$\zeta^{(0)*}$&
$g_0(\nu)\frac{5}{12}(1-e_p^2)(1+\frac{3}{32}c^*(e_p))$
\\
$\gamma^*$&$
\frac{128}{5\pi}\nu^2g_0(\nu)(1+e_p)(1-\frac{1}{32}c^*(e_p))
$\\
$c^*(e_p)$&$32(1-e_p)(1-2e_p^2)[81-17e_p+30e_p^2(1-e_p)]^{-1}.
$\\
\hline
\end{tabular}
\end{table}

The constitutive relations for the 
hydrodynamic equations have been obtained
in ref.~\cite{Garzo} by the Chapman-Esnkog method
with the approximation (\ref{eq:Enskog})
up to the Navier-Stokes order (i.e. the first order of the spatial gradients).
In the simple steady shear flow with constant $n$, constant $T$, and
$\bm u(\bm r)=(\dot \gamma z, 0,0)$,
the nonzero terms are the pressure, or the normal stress
\begin{equation}
N_\alpha\equiv\Sigma_{\alpha,\alpha}=
N(\nu,T)=\sigma^{-3}f_1(\nu)T,
\label{pressure}
\end{equation}
the shear stress
\begin{equation}
S\equiv \Sigma_{x,z}=\Sigma_{z,x}=
S(\nu,T)=m^{1/2}\sigma^{-2}f_2(\nu)\sqrt{T}\dot \gamma,
\label{shear}
\end{equation}
and the energy dissipation rate
\begin{equation}
\Gamma=\Gamma(\nu,T)=m^{-1/2}{\sigma^{-4}}
f_3(\nu)
T^{3/2}.
\label{dissipation}
\end{equation}
The dimensionless functions 
$f_i(\nu)$ are listed in Table~\ref{constitutiveeq}.

In the simple shear flow, 
Eqs.~(\ref{contituity}) and (\ref{motion})
are automatically satisfied with 
the constant normal stress $N$ and the 
constant shear stress $S$.
The energy balance equation (\ref{ebalance0}) gives
\begin{equation}
S\dot \gamma-\Gamma=0,
\label{ebalance}
\end{equation}
because there is no heat flux ${\bm q}$.
Equation~(\ref{ebalance}) means that the granular temperature
is locally determined by the balance between 
the viscous heating and the energy dissipation.
Equation (\ref{ebalance})
with Eqs.~(\ref{shear}) and (\ref{dissipation}) gives
\begin{equation}
T=m\sigma^2\frac{f_2(\nu)}{f_3(\nu)}\dot\gamma^2.
\label{temperaturegamma}
\end{equation}
Substituting Eq.~(\ref{temperaturegamma}) into 
Eqs.~(\ref{pressure}) and (\ref{shear}),
we get
\begin{eqnarray}
N&=& m\sigma^{-1} \frac{f_1(\nu)f_2(\nu)}{f_3(\nu)}\dot \gamma^2,\\
S&=&m\sigma^{-1}\frac{[f_2(\nu)]^{3/2}}{[f_3(\nu)]^{1/2}}\dot \gamma^2,
\end{eqnarray}
which are exactly what we have anticipated from 
the Bagnold scaling Eqs.~(\ref{bagnolds}) and (\ref{bagnoldp}).

The above derivation of the Bagnold scaling by the kinetic theory 
gives the definite expression for Eq.~(\ref{slope}),
\begin{equation}
\frac{S}{N}=\frac{\sqrt{f_2(\nu)f_3(\nu)}}{f_1(\nu)},
\label{SoverP}
\end{equation}
as a function of the packing fraction $\nu$.
This is plotted in Fig.~\ref{friction_packing1}
by symbols connected by lines,
along with the simulation data.
One can see clear discrepancy between the 
theory and the simulation especially in the 
higher density region. The kinetic theory 
gives increasing functions $S/N$ of 
$\nu$, which means that the flow 
down steeper slope is denser.

\section{Simulations}\label{simulation}
In this section, we compare the expressions
Eqs.~(\ref{pressure})-(\ref{dissipation}) 
with the simulation results of simple shear flow
of inelastic hard spheres. 
\subsection{Simulation setup}
The simulation is done under the constant volume condition
with a uniform shear in a rectangular box of the size 
$L_x\times L_y\times L_z$.
The shear is applied by the Lees-Edwards shearing periodic 
boundary conditions in the $z$ direction~\cite{Lees-Edwards};
The periodic boundary condition is employed in the 
$x$ and $y$ directions. 
We employ the event driven method,
using the fast algorithm developed by Isobe~\cite{isobe}.

A steady shear flow with the mean velocity 
$\bm u(\bm r)=(\dot \gamma z,0,0)$ is prepared
as follows. First, a random configuration 
is prepared by the compressing procedure 
proposed by Lubachevsky and Stillinger~\cite{Lubachevski90}
in the elastic system without shear under the 
periodic boundary condition. Secondly, 
the initial shear flow is constructed from the 
above random configuration by giving the initial mean 
velocity $\bm u(\bm r)=(\dot \gamma z,0,0)$ and setting the
initial temperature $T\approx 100m\sigma^2\dot \gamma^2$.
Lastly, the steady shear flow of the 
inelastic system is obtained by relaxing the 
initial flow under the Lees-Edwards shearing periodic 
boundary condition~\cite{comment3}.

With the present parameter and system size, 
the final steady state is the simple shear flow 
with uniform packing fraction $\nu=\nu_0$
and mean velocity $\bm u=(\dot \gamma z,0,0)$~\cite{commentS}.
All the following data are taken in the steady state,
and averaged over the space and time (typically over 10,000
collisions per particle) unless otherwise noted.

In the following, all the quantities are
given in the dimensionless form
with the unit mass $m$, the unit length $\sigma$,
and the unit time $\dot \gamma^{-1}$.
Most of the data are from the simulations with
system size $L_x=20$, $L_y=10$, and $L_z=40$.
Several simulations has been done with $L_x=L_y=L_z=40$ 
to check the system size effect.
We measure the temperature $T$,
the normal stress $N$, the shear stress $S$, and the energy dissipation 
rate $\Gamma$ for various values of the packing fraction $\nu$.
These are compared with Eqs.(\ref{pressure})-(\ref{dissipation})
from the kinetic theory.

\subsection{Simulation results}
\subsubsection{The radial distribution function} \label{radial}
\begin{figure}[tp]
\includegraphics[width=0.46\textwidth]{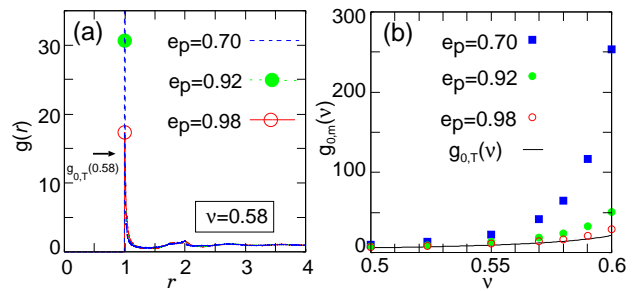}
\caption{(color online)(a)The radial distribution functions 
for $\nu=0.58$ with $e_p=0.7$ (dashed line), 
$0.92$ (dotted line),
and $0.98$ (solid line).
The peak values of contact are 136 (out of range),
30.6 ($\bullet$), and 17.4 ($\circ$) 
for $e_p$=0.70, 0.92, and 0.98, respectively.
The theoretical value at contact,
$g_{0,{\rm T}}(0.58)$, is shown by an arrow.
(b) Plot of $g_{0,{\rm m}}(\nu)$ versus the packing fraction $\nu$
for $e_p=0.7$ ($\blacksquare$), $0.92$ ($\bullet$),
and $0.98$ ($\circ$).
$g_{0,{\rm T}}(\nu)$ is shown by a solid line. 
}\label{gr-plot}
\end{figure}
For the constitutive relations 
with Table~\ref{constitutiveeq}, we need to know 
the radial distribution function at the particle diameter, $g_0(\nu)$,
as a function of the packing fraction.
For elastic hard spheres ($e_p=1$) in equilibrium, 
the well known expression of $g_0(\nu)$
is the Carnahan-Starling formula~\cite{SimpleLiquid}
\begin{equation}
g_{0,{\rm CS}}(\nu)=\frac{1-\nu/2}{(1-\nu)^3}
\end{equation}
for $0<\nu<\nu_f$, where $\nu_f$
is the freezing packing fraction 
and $\nu_f\approx 0.49$  \cite{Torquato}.
Torquato \cite{Torquato} proposed the formula 
that include the higher packing fraction 
up to the random closed packing fraction $\nu_c\approx 0.64$ as
\begin{equation}
g_{0,{\rm T}}(\nu)=
\begin{cases}
g_{0,{\rm CS}}(\nu)&
\mbox{for}\quad 0<\nu<\nu_f,\\
g_{0,{\rm CS}}(\nu_f)(\nu_f-\nu_c)/(\nu-\nu_c)&
\mbox{for}\quad \nu_f<\nu<\nu_c.
\label{Torq}
\end{cases}
\end{equation}

As for the inelastic hard spheres under shear, 
a generally accepted form of $g_0(\nu)$ does not exist,
but it has been found in several simulations that 
$g_0(\nu)$ is larger for stronger inelasticity 
\cite{Bizon,Alam}.
Figure~\ref{gr-plot}(a) shows the radial distribution
$g(r)$ averaged over the all directions
obtained from our shear flow simulation
with the packing fraction $\nu=0.58$ 
for various values of $e_p$.
The spatial mesh to measure $g(r)$ was
taken as $0.001$, and 
the peak values of $g(r)$ around $r=1$
(at the distance of the particle diameter)
are marked by symbols for $e_p=0.98$ and $0.92$.
We can see that the peak value strongly increases
for smaller $e_p$, and can be much larger than the value
from Eq.~(\ref{Torq}) ($g_{0,{\rm T}}(0.58)$ shown by an arrow).
It is quite difficult to evaluate 
the precise value of $g_0(\nu)$ from this direct measurement of $g(r)$
because of the strong increase of $g(r)$ in the limit of $r\to +1$. 

The way we determine $g_0(\nu)$ from the simulation is
through the expression of the collision frequency $\omega_0$~\cite{Lois,Luding}
from the kinetic theory~\cite{Garzo,comment4},
\begin{equation}
\omega_0(\nu,T)=24 g_0(\nu)\sqrt{T}\nu \pi^{-1/2}
\left(1-\frac{1}{32}c^*(e_p)\right),
\label{omega0}
\end{equation}
where $c^*(e_p)$ is given in Table~\ref{constitutiveeq}.
By measuring $\omega_0$ and $T$ 
for each $\nu$ from the simulation,
we can evaluate
\begin{equation}
g_{0,{\rm m}}(\nu;T,\omega_0)
\equiv \frac{\omega_0\sqrt{\pi}}{24(1-c^*(e_p)/32)\sqrt{T}\nu}.
\label{g0m}
\end{equation}
$g_{0,{\rm m}}(\nu;T,\omega_0)$ is plotted versus $\nu$ for various values of $e_p$
in Fig.~\ref{gr-plot}(b), where $g_{0,{\rm T}}(\nu)$ in Eq.~(\ref{Torq}) 
is shown by a solid line 
for reference.
$g_{0,{\rm m}}(\nu;T,\omega_0)$ 
shows stronger increase upon increasing 
the packing fraction $\nu$ as $e_p$ gets smaller; 
by comparing it with Fig.~\ref{gr-plot}(a),
we see that this indirect estimate 
gives an reasonable $e_p$ dependence of $g_0(\nu)$.
In the following, we use
$g_{0,{\rm m}}(\nu;T,\omega_0)$ as $g_0(\nu)$ in Table~\ref{constitutiveeq}
unless otherwise noted.
\subsubsection{The energy dissipation rate and the normal stress
as functions of the packing fraction
}
\begin{figure}[tp]
\includegraphics[width=0.23\textwidth]{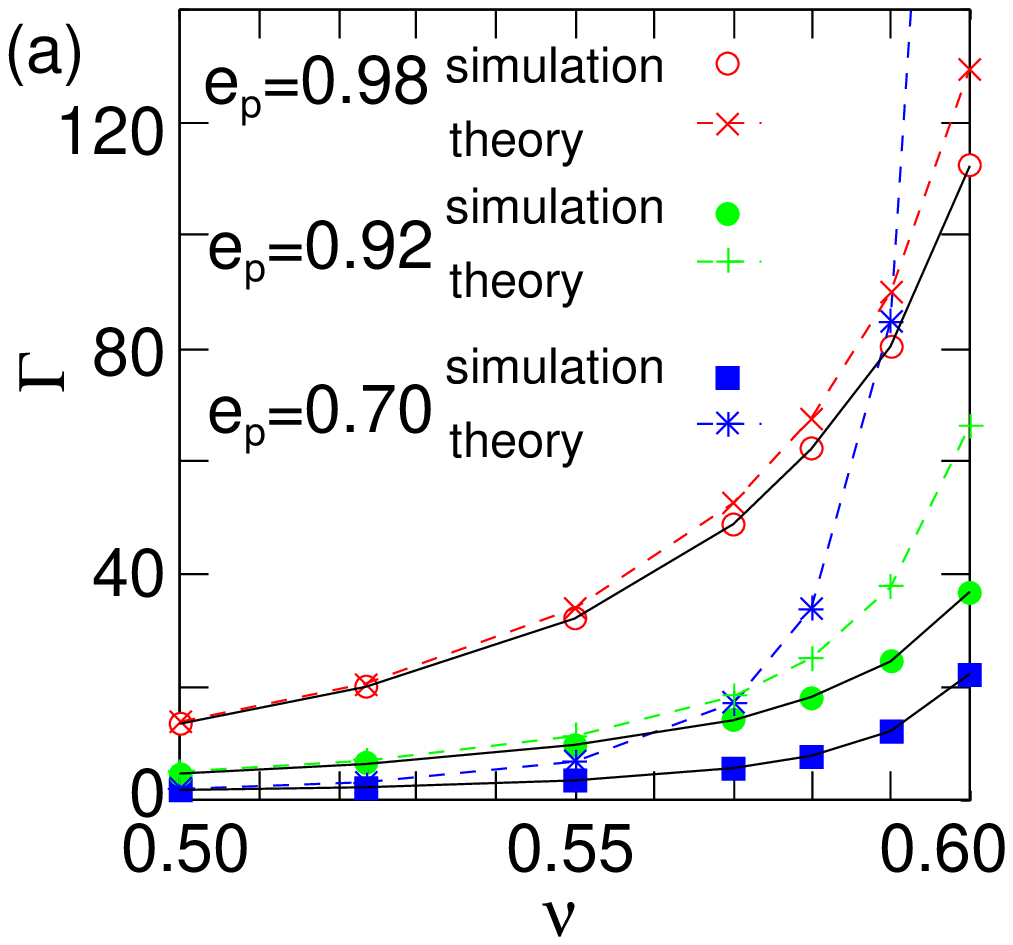}
\includegraphics[width=0.23\textwidth]{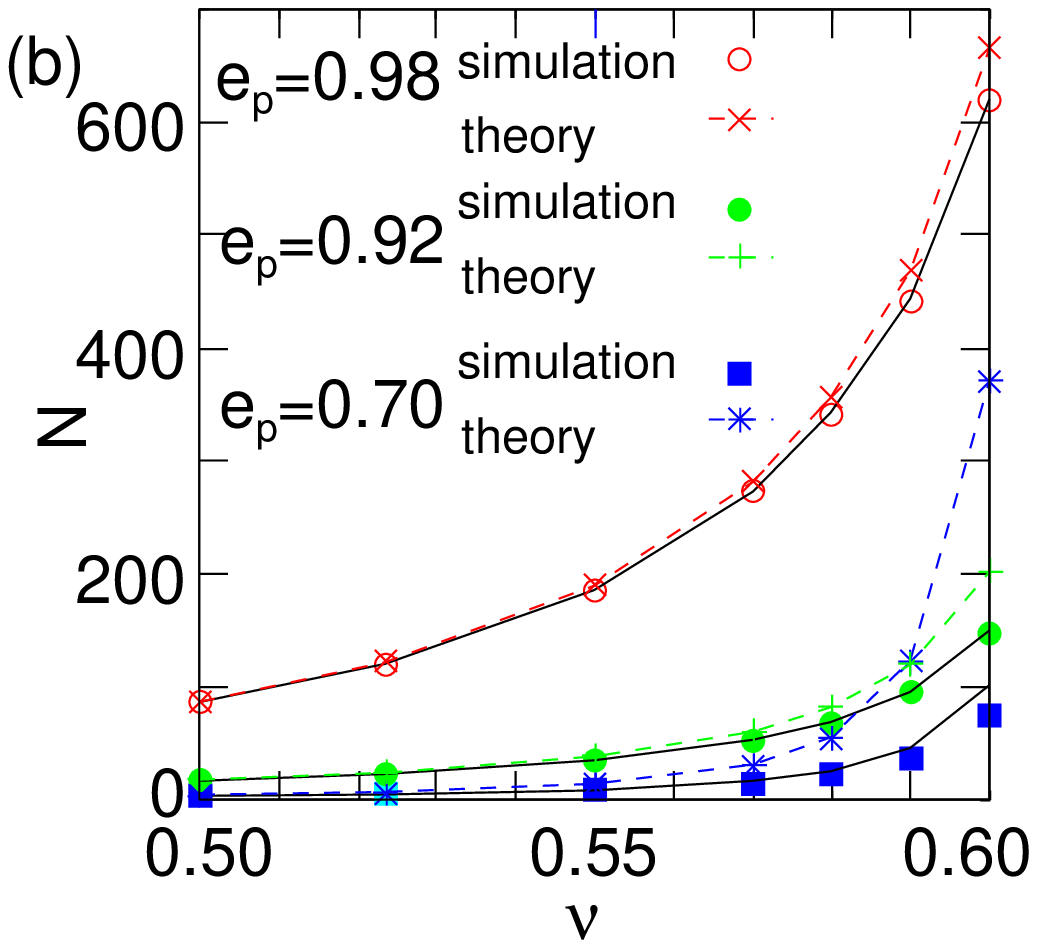}
\caption{(color online)
The energy dissipation rate $\Gamma$ (a)
and the normal stress $N$ (b).
The simulation data for
$e_p=0.98$ ($\circ$), $0.92$ ($\bullet$),
and $0.70$ ($\blacksquare$) are compared with 
the values from the kinetic theory 
($\Gamma(\nu,T)$ and $N(\nu,T)$) shown by 
symbols connected by dashed lines for 
$e_p=0.98$ ($\times$), $0.92$ (+), and $0.70$ ($*$).
$\Gamma(\nu,T_{\rm coll})$ and $N(\nu,T_{\rm coll})$
with $g_{0,{\rm m}}(\nu;T_{\rm coll},\omega_0)$ are denoted by 
the solid lines, which agree with the simulation data (see text).
}\label{Constitutive}
\end{figure}

In Fig.~\ref{Constitutive}, 
the energy dissipation rate $\Gamma$ (a) and the normal stress $N$ (b)
are shown for various values of the 
packing fraction $\nu$ and the restitution coefficient
$e_p$.
For the normal stress, we find in the simulation that $N_\alpha$
depends on the direction $\alpha$, but
the differences among them are
at most $10\%$ in the plotted region
and are not significant compared to the difference from
the kinetic theory that we will study in the following. 
Thus, here we plot the average $N\equiv (N_x+N_y+N_z)/3$.

The value from the kinetic theory are 
shown in Fig.~\ref{Constitutive}(a) and (b)
by symbols connected by dashed lines.
We see in Fig.~\ref{Constitutive}(a)
that the energy dissipation rate is 
overestimated by the theory in the dense region,
and the disagreement is larger for smaller $e_p$.
The normal stress in Fig.~\ref{Constitutive}(b) also shows a similar tendency, 
although the relative disagreements are smaller than 
those in the energy dissipation rate $\Gamma$.  

\subsubsection{The pre-collisional 
velocity correlation effects and the collisional temperature
}\label{precol}
\paragraph{The energy dissipation.}
We first focus on the discrepancy in the 
energy dissipation rate $\Gamma$.
From the collision rule Eq.~(\ref{ColRule}),
the energy dissipated per collision is given by
\begin{equation}
\Delta E_{ij}=(1-e_p^2)\frac{1}{4} c_{n,ij}^2,
\end{equation}
where $c_{n,ij}\equiv \left[\left( \bm{c}_i-\bm{c}_j \right)\cdot 
\bm n_{ij}\right]$ is the relative normal velocity of 
colliding particles just before the collision.
Thus, $\Gamma$ is given by
\begin{equation}
\Gamma=<\Delta E_{ij}>_{\rm coll}\cdot \frac{1}{2}n\omega_0=
(1-e_p^2)\cdot \frac{1}{4} <c_{n,ij}^2>_{\rm coll}
\cdot \frac{1}{2} n\omega_0.
\label{eq:GammaEx}
\end{equation}
Here, $<A>_{\rm coll}$ denotes 
the average of a quantity $A$ over all collisions;
if the value of $A$ is $A_k$ at the $k$-th collision,
$<A>_{\rm coll}\equiv \sum_{k=1}^{N_{\rm coll}} A_k/N_{\rm coll}$,
where $N_{\rm coll}$ is the total number of collisions.
Note that Eq.~(\ref{eq:GammaEx}) is the exact expression for $\Gamma$.

On the other hand, the expression~(\ref{dissipation}) 
from the kinetic theory 
with Eq.~(\ref{g0m}) gives
\begin{equation}
\Gamma(\nu,T)=
(1-e_p^2)\cdot T
\left[
\frac{1+3c^*(e_p)/32}{1-c^*(e_p)/32}
\right]\cdot 
\frac{1}{2} n\omega_0(\nu,T).
\label{gammakinetic}
\end{equation}
To interpret this expression, let us consider
the random collision of particles whose 
velocity fluctuation is given by the Maxwellian. 
In this case, $\frac{1}{4}<c_{n,ij}^2>_{\rm coll}=T$,
then Eq.~(\ref{eq:GammaEx}) gives  
\begin{equation}
\Gamma=(1-e_p^2)T\frac{1}{2}n\omega_0.
\end{equation}
The difference 
between this and Eq.~(\ref{gammakinetic}) 
comes from the deviation of the velocity distribution from the 
Maxwellian, but the difference is
found to be small in the parameter region 
studied in the present paper.
Therefore, from the comparison of the exact expression~(\ref{eq:GammaEx})
with the kinetic theory expression Eq.~(\ref{gammakinetic}),
we conclude that the deviation found in Fig.~\ref{Constitutive}(a)
comes from the fact that 
$\frac{1}{4}<c_{n,ij}^2>_{\rm coll}\ll T$.
\paragraph{The collisional temperature.}
\begin{figure}[tp]
\includegraphics[width=0.23\textwidth]{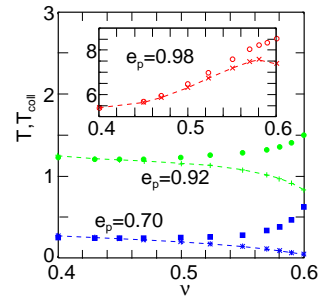}
\caption{(color online)
The temperature $T$ and ``the collisional temperature'' 
$T_{\rm coll}$ vs. the packing fraction $\nu$.
$T$ and $T_{\rm coll}$ are denoted by
$\blacksquare$ and $*$ with the dashed lines
for $e_p=0.7$, respectively, 
and by $\bullet$ and $+$ with the dashed lines
for $e_p=0.92$.
The inset shows $T$ and $T_{\rm coll}$
for $e_p=0.98$ represented by 
$\circ$ and  $\times$ with the dashed lines.}
\label{temperature}
\end{figure}
To confirm this idea, we define ``the collisional temperature''
$T_{\rm coll}\equiv  <c_{n,ij}^2>_{\rm coll}/4$.
Figure~\ref{temperature} shows $T_{\rm coll}$ and $T$
as functions of $\nu$. One can see that $T_{\rm coll}$
is substantially smaller than $T$ for $\nu>0.5$ as is concluded above.

To demonstrate that the discrepancy is 
actually resolved by $T_{\rm coll}$, 
we plot $\Gamma(\nu, T)$
of Eq.~(\ref{dissipation}) with $g_{0,{\rm m}}(\nu, T,\omega_0)$
of (\ref{g0m}) in $f_3(\nu)$ replacing 
$T$ by $T_{\rm coll}$ (the solid lines 
in Fig.\ref{Constitutive}(a));
This is equivalent to 
replace $T$ in Eq.~(\ref{gammakinetic}) with $T_{\rm coll}$ 
and use the measured value of the collision frequency for $\omega_0$.
The agreement is quite good.

\paragraph{Normal stress.}
Now, we consider the effect of $T_{\rm coll}<T$
on the normal stress $N$.
The value of $c_{n,ij}$
should also play an important role 
in the collisional component of the normal stress $N_{\rm coll}$,
because $c_{n,ij}$ directly determines the momentum transfer
from the particle $i$ to the particle $j$ through a collision:
$\Delta \bm p_j
\equiv (\bm c_j^*-\bm c_j)
= [(1+e_p)/2]\  c_{n,ij}\bm n_{ij}=-\Delta \bm p_i$.
Thus, we expect that $N_{\rm coll}$
is approximately 
proportional to $<|\Delta \bm p|>
n\omega_0 \propto \sqrt{T_{\rm coll}} n\omega_0$.
In addition, the collisional part $N_{\rm coll}$
is dominant in the dense region.

In Fig.~\ref{Constitutive}(b), $N(\nu,T_{\rm coll})$ 
of Eq.~(\ref{pressure}) is plotted by the solid lines, where
$g_{0,{\rm m}}(\nu;T_{\rm coll},\omega_0)$ is used as $g_0(\nu)$.
We see that the solid lines
show reasonably good agreement with the data for the whole density
region.

\subsubsection{Origin of the pre-collisional velocity correlation}
\begin{figure}[tp]
\begin{center}
\includegraphics[width=0.23\textwidth]{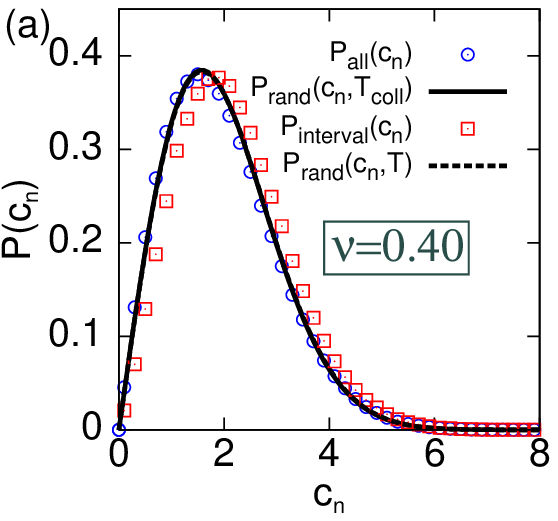}
\includegraphics[width=0.23\textwidth]{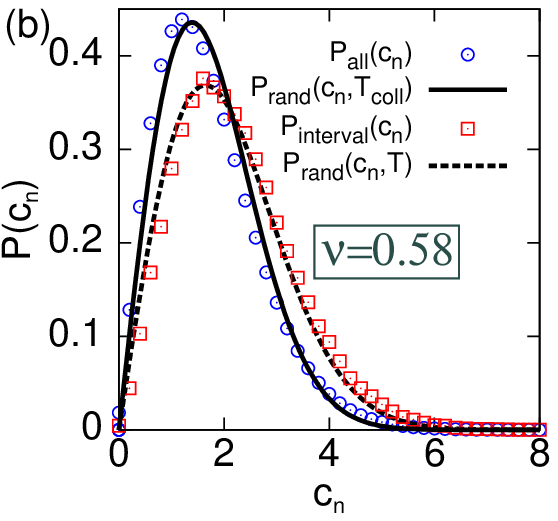}
\end{center}
\caption{(color online)
The distribution of relative normal velocity 
just before the collision $c_n$.
$P_{\rm all}(c_n)$ and
$P_{\rm interval}(c_n)$ are compared along with
$P_{\rm rand}(c_n;T_{\rm coll})$ and $P_{\rm rand}(c_n;T)$ for
$e_p=0.92$ with $\nu=0.40$ (a) and $\nu=0.58$ (b).
We see that the difference between 
$P_{\rm all}(c_n)$ ($\circ$)
and $P_{\rm interval}(c_n)$ ($\square$) is small for
$\nu=0.40$, but for $\nu=0.58$, the $P_{\rm all}(c_n)$ has
sharper distribution.
The solid lines show $P_{\rm rand}(c_n;T_{\rm coll})$ 
and the dashed lines show $P_{\rm rand}(c_n;T)$. 
See text for details.
}\label{colfreq}
\end{figure}
\begin{figure}[tp]
\begin{center}
\includegraphics[width=0.23\textwidth]{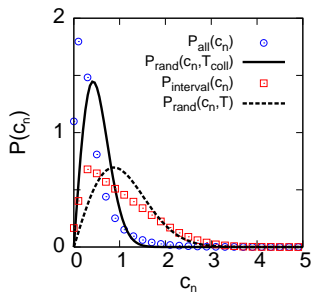}
\end{center}
\caption{(color online)
The pre-collisional relative normal velocity 
distributions 
$P_{\rm all}(c_n)$ ($\circ$) and
$P_{\rm interval}(c_n)$ ($\square$) compared along with
$P_{\rm rand}(c_n;T_{\rm coll})$ (solid line)
and $P_{\rm rand}(c_n;T)$ (dashed line),
respectively, for $e_p=0.70$ with $\nu=0.58$.
}\label{colfreq7}
\end{figure}

One of the possible origins of the 
pre-collisional velocity correlation 
that makes $T_{\rm coll}<T$ 
is the inelasticity,
which makes the relative normal velocity smaller upon collision.
In this subsection, we examine how the pre-collisional velocity 
correlation develops in the shear flow.

It is expected that the correlation grows 
when particles collide with
same colliding partners inelastically many times within a short period of time.
Under the shear, however, this correlation  
will be lost when they are forced to pass each other 
and collide with new partners.
The typical time scale that a pair of particles
pass each other is the unit time, 
i.e., $\dot \gamma^{-1}$.
This argument explains the smaller $T_{\rm coll}$
in the denser region,
because particles collide more frequently
with same partners
before they move far apart~\cite{comment5}.

This argument tells that the collision 
does not have memories of the previous collisions
earlier than the unit time $\dot \gamma^{-1}$. 
To confirm this, we compare 
the following two distributions of the pre-collisional velocity:
(i)$P_{\rm all}(c_{n})$, which is
the distribution of 
$c_{n,ij}$ for all collisions between all pairs of particles,
and (ii) $P_{\rm interval}(c_{n})$, which is
the distribution of $c_{n,ij}$ 
of the collisions whose colliding pairs of particles 
did not collide with each other during the last 
unit time $\dot \gamma^{-1}$.
If the velocity correlation mainly comes from the 
multiple collision with same partners 
within the unit time scale, then $P_{\rm interval}(c_n)$
should have the width determined 
not by $T_{\rm coll}$ but by the average temperature 
$T$. 

The results are shown in Fig.~\ref{colfreq}
for $e_p=0.92$ with $\nu=0.40$~(a)
and $0.58$~(b), where 
$P_{\rm all}(c_{n})$ is denoted by $\circ$ and 
$P_{\rm interval} (c_{n})$ is denoted by $\square$.
We see that $P_{\rm interval}(c_{n})$
is wider than $P_{\rm all}(c_{n})$ for the denser case (b).

If the particles with the Maxwellian velocity 
distribution with temperature $\tilde T$ collide
among themselves randomly,
the distribution of $c_n$ is given by
\begin{equation}
P_{\rm rand}
(c_n;\tilde T)=\frac{c_n}{2 \tilde T}\exp\left[-\frac{c_n^2}{4\tilde T}
\right].
\label{prand}
\end{equation}
In Fig.~\ref{colfreq}, $P_{\rm rand}(c_n;T)$ and 
$P_{\rm rand}(c_n;T_{\rm coll})$ 
are shown by the dashed and solid lines, respectively.
They are indistinguishable for 
$\nu=0.40$ in Fig.~\ref{colfreq}(a),
but show clear difference for $\nu=0.58$ in Fig.~\ref{colfreq}(b).
We find that $P_{\rm rand}(c_n;T)$ fits $P_{\rm interval} (c_{n})$,
and $P_{\rm rand}(c_n;T_{\rm coll})$ 
fits $P_{\rm all} (c_{n})$,
which further confirms that colliding partners 
are correlated in the way characterized by $T_{\rm coll}$.

For smaller $e_p$, the shape of the distributions 
deviates from Eq.~(\ref{prand}) based on the random collision.
Figure~\ref{colfreq7} shows
$P_{\rm all}(c_n)$ and $P_{\rm interval}(c_n)$ compared along with
$P_{\rm rand}(c_n;T_{\rm coll})$ and $P_{\rm rand}(c_n;T)$,
respectively, for $e_p=0.70$ with $\nu=0.58$.
$P_{\rm all}(c_n)$ has sharper distribution than 
$P_{\rm interval}(c_n)$, 
but neither of them fit well with $P_{\rm rand}(c_n;T_{\rm coll})$ 
nor $P_{\rm rand}(c_n;T)$. 
This suggests stronger correlation than the case of $e_p=0.92$.

\subsubsection{The spatial correlation in the velocity fluctuation}
\begin{figure}[tp]
\begin{center}
\includegraphics[width=0.23\textwidth]{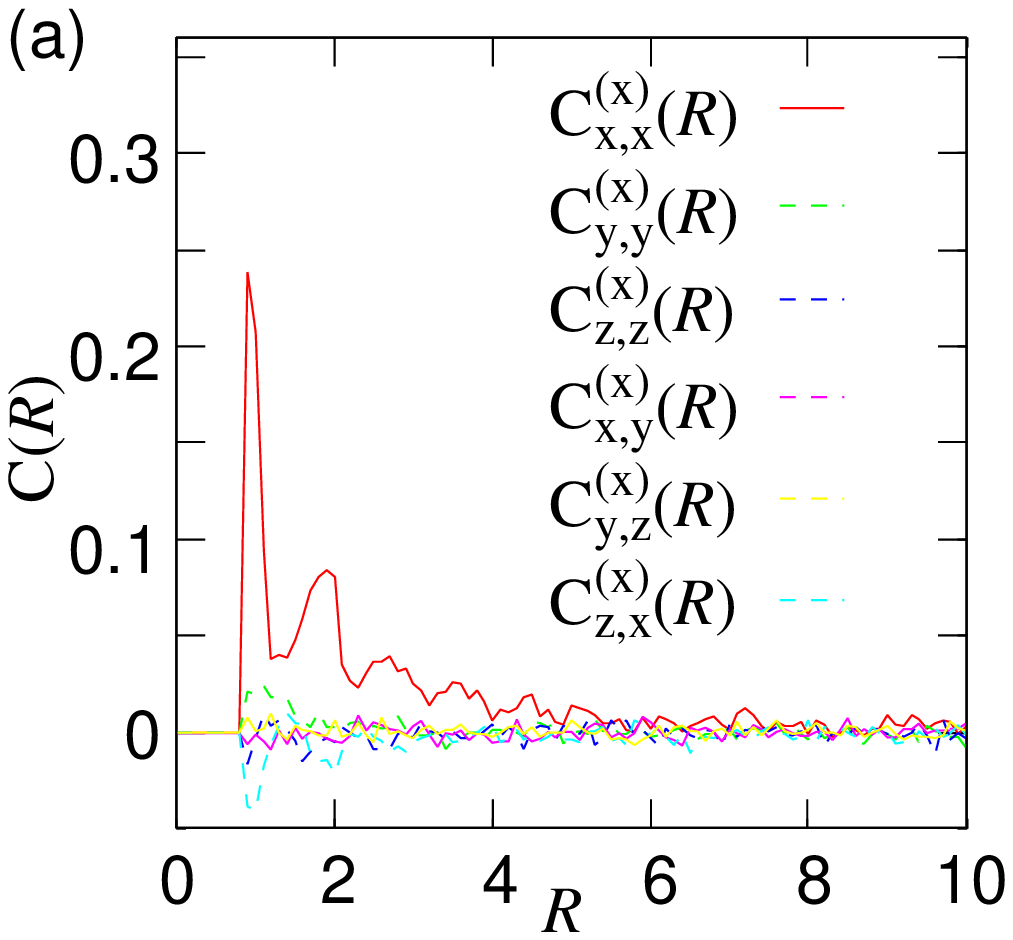}
\includegraphics[width=0.23\textwidth]{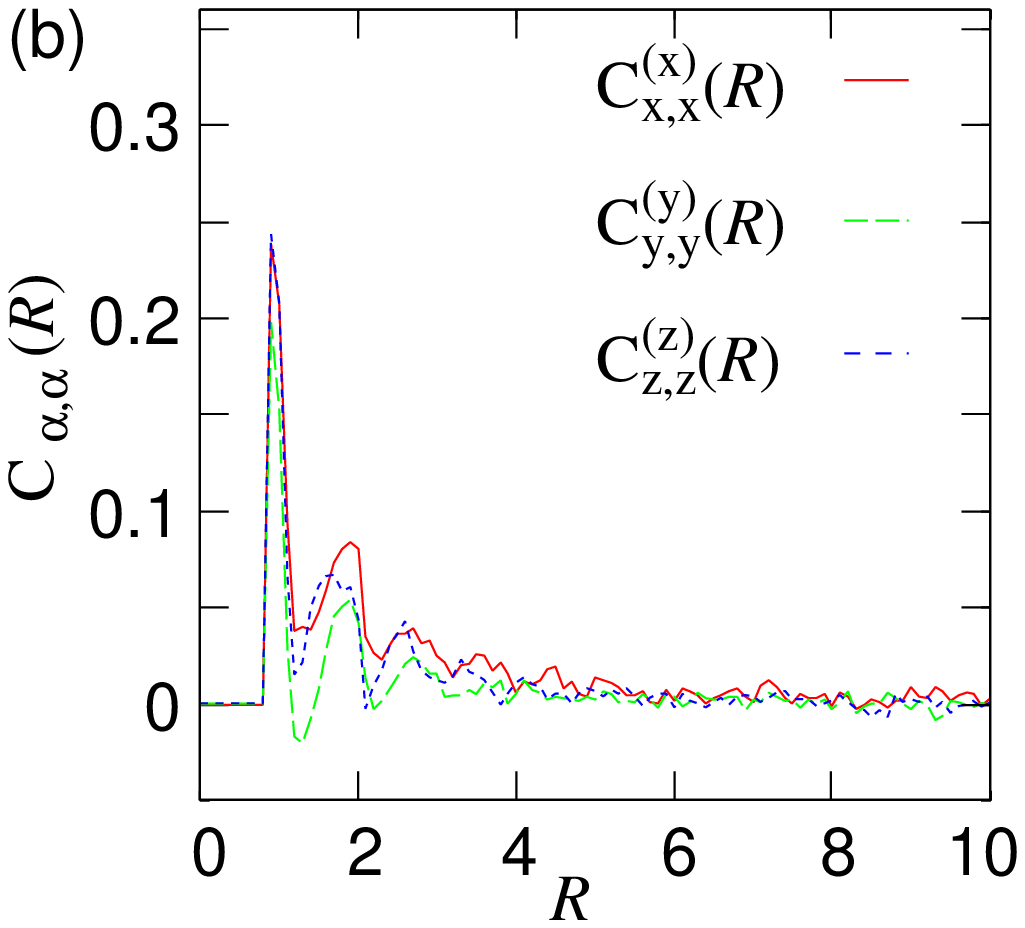}
\end{center}
\caption{(color online)
The spatial velocity correlation functions for
$e_p=0.92$ and $\nu=0.55$.
(a) The correlations in the $x$ direction
$C^{(x)}_{\alpha,\beta}(R)$. One sees that
the longitudinal component
$C^{(x)}_{x,x}(R)$ (solid line) has the larger amplitude
than others (shown by dashed lines).
(b) The longitudinal velocity correlations
$C^{\alpha}_{\alpha,\alpha}(R)$
for $\alpha=x$ (solid line),
$y$ (dashed line), and $z$ (dotted line).
}\label{correlation}
\end{figure}

To understand the the velocity correlations
in more detail,
we study the 
spatial velocity correlation function defined as
\begin{equation}
C^{(\gamma)}_{\alpha,\beta}(R)=
\frac{<\sum_{i,j}\left[\tilde c_{\alpha,i}\tilde c_{\beta,j}
\Delta(R-|(\bm r_i-\bm r_j)\cdot \bm e_\gamma|) 
\right]>
}{<\sum_{i,j}\left[
\Delta(R-|(\bm r_i-\bm r_j)\cdot \bm e_\gamma| )
\right]>},
\end{equation}
where $\alpha$, $\beta$, and $\gamma$
take $x,y$ or $z$, 
$\tilde c_{\alpha,i}\equiv(c_{\alpha,i}-u_\alpha)$, 
$<\cdots>$ denotes the time average,
and $\bm e_\gamma$  represents the unit vector 
in the $\gamma$ direction. 
$\Delta(R-|\bm r\cdot \bm e_\gamma| )$
is one when
$|R-|\bm r\cdot \bm e_\gamma||<0.05$ and
$|\bm r\cdot \bm e_{\gamma'}|<0.1$ for $\gamma'\ne \gamma$,
and it is zero otherwise.
We calculated $C^{(\gamma)}_{\alpha,\beta}(R)$ for the system
with $e_p=0.92$, both for the small system with
$L_x=20,L_y=10,L_z=40$  and
the large system with $L_x=40,L_y=40,L_z=40$.
We find that the correlation extends over the whole system
in the case of the small system,
but it goes to zero for the large system.
In the following, we present the spatial
correlation measured in the large system,
but we confirmed that 
the hydrodynamic quantities
presented in the previous subsections
did not show any differences.

In Fig.~\ref{correlation}(a),
the various components of the correlation in 
the $x$-direction $C^{(x)}_{\alpha,\beta}(R)$ are shown.
We find that the longitudinal correlation 
in the $x$-direction,  $C^{(x)}_{x,x}(R)$,
has larger amplitude than other components;
this tendency is also found in the $y$- and $z$-direction (data not shown).
The longitudinal correlation at the particle diameter 
distance ($R=1$) is positive,
which is consistent with the fact that $T_{\rm coll}<T$.
It is evident that the correlation shows an oscillation,
whose wavelength is order of the particle diameter,
which will be discussed in section \ref{discussion}.

The longitudinal components in $x$, $y$ and $z$ directions
are shown in Fig.~\ref{correlation}(b).
All of them show oscillations in the particle diameter scale.
We also found that the longitudinal correlation shows larger 
amplitude for smaller restitution coefficient $e_p$ 
and/or larger packing fraction $\nu$ (data not shown).

\subsubsection{The packing fraction dependence of the 
shear stress}
\begin{figure}[tp]
\includegraphics[width=0.23\textwidth]{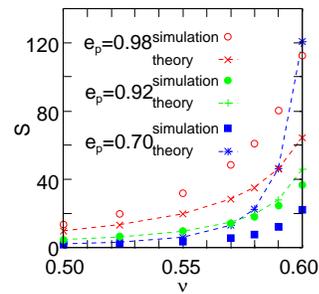}
\caption{(color online) The shear stress $S$ vs. 
the packing fraction $\nu$. 
The simulation data are plotted for
$e_p=0.98$ ($\circ$), $0.92$ ($\bullet$), 
and $0.70$ ($\blacksquare$).
The kinetic theory constitutive relations $S(\nu,T)$
is shown by the symbols connected by dashed lines
($\times$ for $e_p=0.98$, $+$ for $0.92$, and 
$*$ for $e_p=0.70$).
}\label{shearfig}
\end{figure}

We find that the shear stress $S$
shows more complicated packing fraction $\nu$ dependence 
than those of the energy dissipation rate $\Gamma$ and the 
normal stress $N$. 
In Fig.\ref{shearfig}, the simulation data of the shear stress $S$
are denoted by symbols, and $S(\nu,T)$ from the kinetic 
theory (Eq.~(\ref{shear})
with Table~\ref{constitutiveeq}) are denoted by 
symbols with the dashed lines.
We find that,
for $e_p=0.98$, the shear stress is {\it underestimated} by the theory, 
while for $e_p=0.92$  and $0.70$, 
the shear stress is {\it overestimated}.

Actually, in the case of the elastic ($e_p=1$) hard sphere 
system, the Enskog theory is known to underestimate
the shear viscosity in dense region~\cite{SimpleLiquid,Alder},
and this tendency is seen in the result for $e_p=0.98$.
The results for $e_p=0.70$ shows that the inelasticity reduces
the shear stress to the value smaller than 
the one expected from the kinetic theory, but 
we do not understand the reason of this reduction yet.
Rather good agreement in between for the case of $e_p = 0.92$ 
seems to be accidental.

\section{Discussion and summary}\label{discussion}
\subsection{The shear stress and 
the anisotropic correlation}
In contrast to the 
energy dissipation rate $\Gamma$
and the normal stress $N$, 
the discrepancy in the shear stress $S$
cannot be understood 
just by the pre-collisional 
velocity distribution averaged over all directions,
but the anisotropy of the pre-collisional 
correlations in both the velocity and the position
should by important in the shear stress. These 
anisotropies are not taken into account in the kinetic theory
employed in the preset analysis. In fact,
for the soft-sphere system in two-dimensional, sheared flow, 
it has been found that the contact force distribution
strongly depends on direction \cite{Cruz}.
Our preliminary results also show a similar direction 
dependence in the collisional momentum transfer per unit time.
The detailed analysis is left for future studies.

\subsection{The packing fraction dependence of the 
ratio of the shear stress to the normal stress}
As we have seen in Fig.~\ref{friction_packing1},
$S/N$ in the simulation is a decreasing function of the 
packing fraction $\nu$ for larger 
packing fraction $\nu$ and/or smaller 
restitution coefficient $e_p$, while 
Eq.~(\ref{SoverP}), from the kinetic theory, 
$S/N$ always increases with $\nu$.

Kumaran argued that the particle roughness 
is necessary for $S/N$ to have a decreasing part
upon increasing $\nu$ in the dense region~\cite{Kumaran}.
However, even for the smooth particles,
the present simulations show that
$S/N$ has a decreasing part in the dense region
for the inelastic hard sphere system,
although the particle roughness may
well amplify the decreasing part of $S/N$.

The present authors have suggested~\cite{Mitarai}
that the origin that leads the kinetic theory to the increasing
$S/N$ on $\nu$ even for the denser region is that
$f_3(\nu)$ in the energy dissipation $\Gamma$
of Eq.~(\ref{dissipation}) increases too sharply for larger $\nu$.
In this paper in section~\ref{precol}, we showed that
the sharp increase in $\Gamma$ can be weaken by 
using $T_{\rm coll}$ instead of $T$.
In the present treatment, however,
It is not possible to extract the 
$\nu$-dependence out of $\Gamma(\nu, T_{\rm coll})$
and to compare it directly 
with $f_3(\nu)$ because $T$ and $T_{\rm coll}$
are determined by $\nu$ and $\dot \gamma$
in the steady state simulations, therefore, the 
quantity that corresponds to $f_3(\nu)$ in 
eq.~(\ref{dissipation}) cannot be defined from the simulation data.

Finally, let us comment on
the fact that $S/N$ does increase with $\nu$ 
in a certain parameter range in our simple shear flow simulation, 
in contrast to the fact that the increasing 
$\nu$ upon increasing $S/N=\tan\theta$
has never been observed in the granular flow down a slope.
This suggests that the steady flow in this parameter region is 
unstable in the slope flow configuration.
It is interesting to study 
the relation between the stability of the flow and 
the $\nu$ dependence of $S/N$.

\subsection{Oscillation in the spatial velocity correlation}
As shown in Fig.~\ref{correlation}, the spatial 
velocity correlation is found to oscillate in the scale
of the particle diameter.
Although we have not yet understood the origin of 
this oscillation, it is plausible that 
the oscillation comes from 
the coupling between
the density correlation and the velocity correlation.
Analysis on the sheared Langevin system suggests that
the spatial velocity correlation is related to the 
radial distribution function~\cite{Yoshimori}, which oscillates
in the particle diameter scale.
It is likely that similar coupling also exists
in the granular shear flow.

\subsection{Summary}\label{conclude}
We have simulated the simple shear flow of 
the smooth inelastic hard sphere system by molecular dynamics
simulations. 
We have found that the energy dissipation rate $\Gamma$
and the normal stress $N$ are 
smaller than those expected from the kinetic 
theory.  
We have showed that the relative pre-collisional 
normal velocity of colliding pairs 
of particles, $c_{n,ij}$, is smaller than the one expected 
from random collisions, and this reduces $\Gamma$ and $N$.
By examining the distributions of $c_{n,ij}$
for all collisions ($P_{\rm all}(c_n)$) and 
for only the first collisions of the new pairs 
during the last period of time $\dot \gamma^{-1}$
($P_{\rm interval}(c_n)$), we have concluded  
that the reduction of the relative velocity 
is caused by the multiple inelastic collisions 
during the time period $\dot \gamma^{-1}$.

To understand the velocity correlation in more detail,
we have studied the spatial velocity correlation.
It has been found that the longitudinal 
components of the correlations have larger amplitude
with the oscillation in the scale of 
the particle diameter.

The shear stress $S$ 
has been found to be overestimated 
for smaller $e_p$, but underestimated for larger $e_p$ 
by the kinetic theory.

\begin{acknowledgments}
NM thanks A.~Yoshimori for his insightful discussion on the 
spatial velocity correlation in the Langevin system.
NM is supported in part by the Inamori foundation. 
Part of this work has been done
when NM was supported by  Grant-in-Aid for Young 
Scientists(B) 17740262 from The Ministry of Education,
Culture, Sports and Technology (MEXT),
and HN and NM were supported by
Grant-in-Aid for Scientific Research (C) 16540344 
from Japan Society for the Promotion of Science (JSPS).
\end{acknowledgments}



\begin{thebibliography}{9}
\bibitem{comment1}
Experimentally,
the spatial ``velocity correlation''
was studied by Pouliquen~\cite{PouliquenCor}
in the granular flow down a slope. 
In the experiments, however, the velocity 
is measured from displacement between certain time intervals
(limited by camera frames), and it 
is different from the simultaneous velocity 
correlations studied in the present paper.
\bibitem{comment3}
In the case of the small restitution coefficient (e.g. $e_p=0.7$),
the system often encounters the inelastic 
collapse~\cite{Collapse,CollapseShear} before reaching the steady state
unless we start the simulation with the initial mean flow. 
With the initial mean flow, the system may reach
a steady shear flow, but the inelastic collapse may eventually occur
even with shear~\cite{CollapseShear}.
In this paper, we focus on the results of the simulations where
the inelastic collapse did not occur within the run.
\bibitem{commentS}
The simple uniform shear flow can be unstable for 
large system~\cite{stability},
but we do not pursue the condition for its stability.
\bibitem{comment4}
In ref.~\cite{Garzo},
the collision frequency $\omega_0$
was not explicitly calculated,
but the value of the necessary integral was given.
\bibitem{comment5}
In the denser region of the shear flow, 
i.e. with smaller free volume,
the particles tends to move collectively 
and this may also create the velocity correlation 
in the dense regime even without inelasticity.
In any case, the collective motion is 
broken by the shear, and the time scale
that the shear breaks the correlation is 
about the unit time ($\dot \gamma^{-1}$).
\end{thebibliography}
\end{document}